\documentclass{ws-procs9x6}

\begin{document}

\title{Singlet NN Regularization using a Boundary Condition for OPE
and TPE potentials\footnote{\uppercase{W}ork partially supported by the spanish
\uppercase{DGI} grant no. \uppercase{BMF}2002-03218, \uppercase{J}unta
de \uppercase{A}ndaluc\'{\i}a grant no. \uppercase{FM}-225 and
\uppercase{EURIDICE} grant number \uppercase{HPRN-CT}-2003-00311.}}
\author{M. Pav\'on Valderrama~\footnote{\uppercase{S}peaker at
\uppercase{NSTAR} 2004; 24-27 \uppercase{M}ar 2004,
\uppercase{G}renoble, \uppercase{F}rance} \, \, and E. Ruiz Arriola}
\address{Departamento de F\'{\i}sica Moderna, Universidad de Granada,
\\ E-18071 Granada, Spain. \\ E-mail: mpavon@ugr.es and
earriola@ugr.es}

\maketitle

\abstracts{Boundary conditions provide a simple and physically
compelling Wilsonian renormalization method in coordinate space in the
framework of effective field theory as applied to nucleon-nucleon
interaction in the non perturbative regime, which enable to remove
unphysical regularization cut-offs while keeping physical low energy
threshold parameters such as the scattering length and effective range
invariant. We illustrate how the method successfully works in the
$^1S_0$ channel for the One Pion Exchange (OPE) and Two Pion Exchange
(TPE) potentials.}


While it is widely accepted that the NN-interaction is not well known
at short distances and different NN potentials~\cite{Stoks:1993tb}
disagree most below $0.5 {\rm fm}$, there is an unanimous agreement
on the fact that pion exchanges~\cite{Rentmeester:1999vw} govern their
long distance behaviour in the spirit of an Effective Field
Theory~\cite{Bedaque:2002mn}. Unfortunately, practical
calculations~\cite{Rentmeester:1999vw,Beane:2001bc,Entem:2003ft,Epelbaum:2004fk}
require finite cut-offs, violating one-valuedness and renormalization
group invariance~\cite{Birse:1998dk} in the non-perturbative low
partial waves. In this talk we show how cut-offs can be removed in the
simplest singlet $^1S_0 $ channel on the light of our recent
results~\cite{Valderrama:2003np,Valderrama:2004nb} where other
(coupled) channels are also studied.

The Schr\"odinger equation in the $^1S_0$ channel at a given momentum
$k$, corresponding to an asymptotic scattering state reads
\begin{equation}
-u_k '' (r) + U(r) u_k (r) = k^2 u_k (r) \qquad , \, u_k (r) \to \sin (k r
+ \delta(k))   \, , 
\label{eq:sch} 
\end{equation}
where $u_k(r)$ and $U(r)= M_N V(r) $ are the reduced wave function and
potential respectively. Long distance properties are encoded in the
potential $U(r)= U_{\rm OPE}(r) + U_{\rm TPE} (r) + \dots
$~\cite{Rentmeester:1999vw}. On the other hand, for regular
potentials, $U(r) > -1/4r^2 $, the short distance properties can be
described by an energy dependent short distance boundary condition
(BC),
\begin{eqnarray}
u_k^\prime  (0) - k \cot \delta_S (k) u_k (0)=0 
\end{eqnarray}
which can also be written in terms of a ``short distance'' phase shift
$\delta_S (k)$ , as can be seen by setting the long distance
contribution $U(r)$ to zero. Strictly speaking $\delta_S (k)$ is
unknown. In fact, the whole problem has to do with a long distance
distortion of the short distance $\delta_S(k)$ through the pion
exchange potential $U(r)$ to produce the total phase shift
$\delta(k)$. Although this may appear to be a hopeless program at
first sight some information can be obtained via a low energy
expansion of the short distance boundary condition. This is equivalent
to an effective range (ER) expansion of $\delta_S (k)$,
\begin{eqnarray}
k \cot \delta_S (k) = \lim_{R_S \to 0} \frac{u_k' (R_S)}{u_k (R_S)} =
- \frac1{\alpha_S} + \frac12 r_S k^2 + \dots
\label{eq:low}
\end{eqnarray} 
where $\alpha_S$, $r_S$, are the short distance scattering length and
effective range parameters respectively. The limit $R_S \to 0 $ has to
be taken carefully because $U(r)$ is singular at the origin.
Obviously, this expansion can be pursued to any order in the
momentum~\cite{Valderrama:2003np}. The expansion of Eq.~(\ref{eq:low})
carries over to the Schr\"odinger equation, Eq.~(\ref{eq:sch}),
yielding
\begin{eqnarray}
u_k (r)= u_0 (r) + k^2 u_2 (r) + \dots   \, , 
\end{eqnarray} 
where at LO we get  
\begin{eqnarray} 
-u_0 '' (r) + U(r) u_0 (r) &=& 0  \nonumber \\ \alpha_S u_0' (0) +  u_0
 (0) &=& 0 \qquad  \qquad  \\  
u_0 (r) &\to&   1- r/\alpha  \nonumber 
\end{eqnarray} 
and at NLO 
\begin{eqnarray} 
-u_2 '' (r) + U(r) u_2 (r) &=& u_0 (r) \nonumber \\ \alpha_S u_2' (0)
+ u_2 (0) &=& \frac12 r_S \alpha_S u_0 (0) \\ u_2 &\to& r \left(r^2 -3
\alpha r + 3 \alpha r_0 \right) / 6 \alpha \nonumber
\end{eqnarray}
and so on. The standard way to proceed would be to integrate the
equations from the origin (or a sufficiently small radius $R_S \to 0$)
and then to adjust the short distance parameters to get the proper
threshold parameters, $\alpha$ and $r_0$. An additional difficulty
hinders the practical calculations; the strong and attractive
singularities of the long distance potential at short distances
($-C/r^6 $ for TPE ) generate increasingly large oscillations of the
wave function close to the origin, so that a extreme fine tunning of
the BC is needed. In addition, the oscillations of the wave function
enable the usage of semi-classical methods to improve on the accuracy
in the short distance region. 

\begin{figure}[ht]
\centerline{\epsfxsize=3.1in\epsfbox{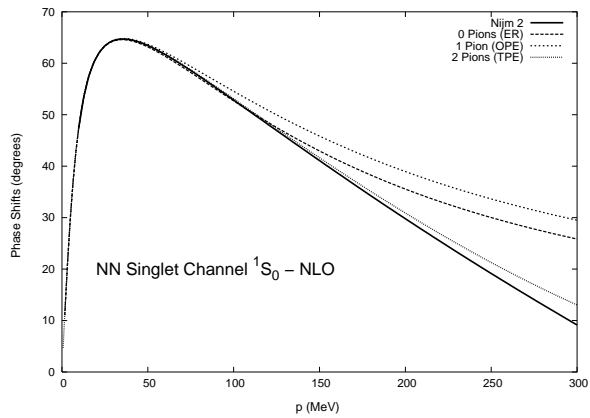}}   
\caption{Renormalized $^1S_0$ phase shift for theory with no pions,
OPE and TPE. In all cases the experimental scattering length and
effective range are taken as input, $ \alpha= -23.73 {\rm fm} $ $ r_0
= 2.67 {\rm fm}$ (see main text) respectively. The short distance
(ultraviolet) cut-off has been completely removed $R_S \to 0$.}
\label{fig:phaseshift}
\end{figure}

Instead of the fitting procedure, one can simply integrate from
infinity downwards, with a known value of $\alpha $, $r_0$ and so on,
to the origin~\cite{Valderrama:2003np,Valderrama:2004nb}. This
provides the value of the short distance parameters, $\alpha_S $ and
$r_S$ for $R_S \to 0$, and then one can use Eq.~(\ref{eq:sch}) to
compute for any energy with a given truncated boundary condition. This
procedure provides by definition the low energy parameters we started
with and takes into account that the long range potential determines
the form of the wave function at long distances. The only parameter in
the procedure is the short distance radius $R_S$ which, unlike
previous work~\cite{Rentmeester:1999vw}, we remove it by taking the
limit $R_S \to 0$. Following this procedure we obtain
Fig.~\ref{fig:phaseshift}, for the theory without pions (ER), as well
as the OPE and TPE potentials. The agreement to
data~\cite{Stoks:1993tb} is good up to the region $ k \sim m_\pi $
which corresponds to the two pion left cut.

\begin{figure}[ht]
\centerline{\epsfxsize=3.1in\epsfbox{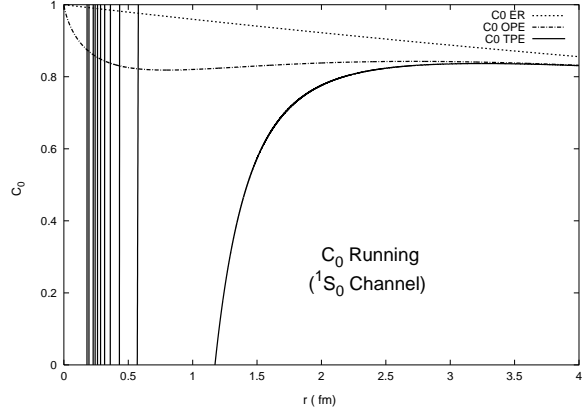}}   
\caption{Evolution of the dimensionless renormalization constant $C_0
(R)$ given by Eq.~(\ref{eq:C0}) for the theory without pions (ER), One
Pion Exchange (OPE) and Two Pion Exchange (TPE). Cycles below 0.2 {\rm
fm} are not plotted. 
\label{fig:C0}}
\end{figure}

Finally, the relation with other renormalization methods, such as the
sharp momentum cut-off method~\cite{Birse:1998dk} used in the Lippmann
Schwinger equation with $ \langle k'| U | k \rangle = - M_N^2 C_0 (\Lambda )
/ 4 \Lambda + \dots $ can be established straightforwardly yielding
\begin{eqnarray} 
C_0 = 1 - R \frac{u_0^\prime (R) }{u_0(R)} \to \frac{\alpha}{\alpha-R}
\qquad , \, R >> 1/m_\pi \, , 
\label{eq:C0} 
\end{eqnarray} 
where $\Lambda = \pi /2 R $. In Fig.~(\ref{fig:C0}) we plot the
dependence of the renormalization constant at a given scale $R$.  As
we see at the OPE level, $C_0 \to 0 $, which corresponds to an
ultraviolet fixed point. This behaviour is changed for TPE to an
oscillating $ C_0 $ with a varying period corresponding to a limit
cycle, illustrating the non-trivial evolution of the BC.

\end{document}